\pdfoutput=1
\documentclass[aps,prb,showpacs,preprintnumbers,twocolumn]{revtex4-1}

\usepackage[colorlinks=true, pdfstartview=FitV, linkcolor=blue, citecolor=blue, urlcolor=blue]{hyperref}
\usepackage[dvipdfmx]{graphicx}
\usepackage{amsmath}
\usepackage{color}

\newcommand{\Z}{\textbf{Z}}

\begin{document}

\title{Quantum Monte Carlo simulation of topological phase transitions}

\author{Arata~Yamamoto}
\affiliation{Department of Physics, The University of Tokyo, Tokyo 113-0033, Japan}

\author{Taro~Kimura}
\affiliation{Department of Physics, Keio University, Kanagawa 223-8521, Japan}

\date{\today}
\begin{abstract}
We study the electron-electron interaction effects on topological phase transitions by the ab-initio quantum Monte Carlo simulation.
We analyze two-dimensional class A topological insulators and three-dimensional Weyl semimetals with the long-range Coulomb interaction.
The direct computation of the Chern number shows the electron-electron interaction modifies or extinguishes topological phase transitions.
\end{abstract}

\maketitle

\section{Introduction}

Study of topological phenomena is one of the most active research areas in condensed-matter physics these days.
The classification of the possible topological phases has been completed within free fermion theory~\cite{Schnyder:2008tya,Kitaev:2009mg}.
However, understanding of the electron correlation effect on such topological phenomena still requires a lot of works in this field.
There have been several proposals for the correlation effect on the topological phases, e.g. breakdown of the free theory classification $\Z \to \Z_{8(16)}$~\cite{Fidkowski:2009dba}, topological Mott insulator~\cite{Pesin:2009NP} and Kondo insulator~\cite{Takimoto:2011JPSJ,Dzero:2012PRB}, fractional topological phases~\cite{Sheng:2011NC,Venderbos:2012PRL}, and so on.

In order to deal with the correlation effect, one often applies the mean-field approximation.
This approximation is not reliable under sufficiently large thermal and quantum fluctuations, typically appearing in the vicinity of phase transitions.
For example, the mean-field analysis shows an artificial phase transition instead of a correct cross-over behavior, e.g. in the Kondo effect.
This implies that we need more reliable methods to deal with the correlation effect, in particular, on the critical phenomena beyond the mean-field approximation.
Furthermore, a typical model applied to study the correlation effect is the Hubbard-type model~\cite{[{See a review article: }] [{ and references therein.}]Hohenadler:2013JPC}.
Since the Hubbard interaction is written as an on-site term, its computational cost is relatively cheap.
However, such a short-range interaction is less realistic in the actual electron system than the long-range Coulomb interaction.
The quantum Monte Carlo method is an effective choice to take into account both of these important pieces.
We remark several quantum Monte Carlo works on graphenes or semimetals with the long-range Coulomb interaction \cite{2011PhRvL.106w6805W,Ulybyshev:2013swa,Smith:2014tha,2014PhRvB..90h5146H,Luu:2015gpl,2016arXiv160807162B}, and also the Chern insulator with the short-range Hubbard interaction~\cite{Hung:2014PRB,Meng:2014MPLB,Wu:2015PRB}.

In this paper, we establish an ab-initio method to investigate topological phase transitions in the presence of the Coulomb-type long-range interaction.
Topological phase transitions are caused by topology change, which cannot be described by conventional order parameters of symmetry breaking.
We adopt the direct calculation of topological invariant using quantum Monte Carlo method~\cite{Yamamoto:2016rfr}.
Based on this approach, we study quantum topological phase transitions of the class A systems both in two and three dimensions involving the long-range interaction.
We show that the quantum critical point is shifted due to the interaction effect, and the region exhibiting the non-trivial topological phase gets small.
We also show that the averaged topological number takes non-integer value in the vicinity of the topological phase transition.
This is interpreted as a consequence of quantum fluctuation around the quantum critical point.

This paper is organized as follows.
In Sec.~\ref{sec2}, we introduce the path-integral formalism to calculate the Berry curvature and the Chern number~\cite{Yamamoto:2016rfr}.
Using this formalism, we analyze two-dimensional class A topological insulators in Sec.~\ref{sec3}, and three-dimensional Weyl semimetals in Sec.~\ref{sec4}.
Finally, we summarize our conclusion in Sec.~\ref{sec5}.

\section{Formalism}
\label{sec2}

Let us consider the path integral with imaginary time $\tau$.
The path integral is given by the Grassmann integral of spin-up electrons and spin-down electrons,
\begin{equation}
Z = \int {\mathcal D}\psi_\uparrow {\mathcal D}\psi^\dagger_\uparrow {\mathcal D}\psi_\downarrow {\mathcal D}\psi^\dagger_\downarrow \, e^{-S} 
.
\end{equation}
We consider the action
\begin{equation}
\begin{split}
S
=& \int d\tau \sum_{x,x'} \Bigg[ \sum_{s=\uparrow,\downarrow} \psi^\dagger_s (x,\tau) K_0(x,\tau|x',\tau) \psi_s(x',\tau)
\\
&+ V(x|x') \psi^\dagger_{\uparrow} (x,\tau) \psi_{\uparrow}(x,\tau) \psi^\dagger_{\downarrow} (x',\tau) \psi_{\downarrow}(x',\tau) \Bigg]
.
\end{split}
\end{equation}
The non-interacting Dirac operator $K_0(x,\tau|x',\tau)$ is given by
\begin{equation}
K_0(x,\tau|x',\tau) = \frac{\partial}{\partial \tau} - H_0(x|x')
,
\end{equation}
where the non-interacting Hamiltonian $H_0(x|x')$.
The matrix $V(x|x')$ controls the interaction between the spin-up electron at $x$ and the spin-down electron at $x'$.
The electron-electron interaction is rewritten by the auxiliary field $\eta(x,\tau)$ through the Hubbard-Stratonovich transformation.
The path integral becomes
\begin{equation}
\label{eqZprime}
Z = \int {\mathcal D}\psi_\uparrow {\mathcal D}\psi^\dagger_\uparrow {\mathcal D}\psi_\downarrow {\mathcal D}\psi^\dagger_\downarrow {\mathcal D}\eta \, e^{-S'} 
\end{equation}
and the action becomes
\begin{equation}
\label{eqSprime}
\begin{split}
S'
=& \int d\tau \sum_{x,x'} \Bigg[ \sum_{s=\uparrow,\downarrow} \psi^\dagger_s (x,\tau) K(x,\tau|x',\tau) \psi_s(x',\tau)
\\
& + \frac{1}{2} V^{-1}(x|x') \eta(x,\tau)  \eta(x',\tau) \Bigg]
\end{split}
\end{equation}
with the interacting Dirac operator
\begin{equation}
K(x,\tau|x',\tau) = \frac{\partial}{\partial \tau} - H_0(x|x') +i\eta(x,\tau)
.
\end{equation}
The electron propagator is given by the inverse Dirac operator $K^{-1}(x,\tau|x',\tau')$.
We remark that the model after the Hubbard-Stratonovich transformation is also interpreted as a system involving the correlated random potential \cite{2015PhRvL.115x6603C,2016PhRvL.116f6401L,2016arXiv160908368L}.

We define the fixed-momentum electron state
\begin{equation}
\begin{split}
\label{eqphipt}
\phi (p,\tau) 
=& \ K^{-1}(p,\tau) \phi_{\rm init}
\\
=& \ \sum_{x,x'} e^{i \sum_j p_j (x_j-x'_j)} K^{-1}(x,\tau|x',0) \phi_{\rm init}
,
\end{split}
\end{equation}
where $\phi_{\rm init}$ is arbitrary initial state.
This state grows or damps in the imaginary-time direction as 
\begin{equation}
 \phi (p,\tau) = \sum_{n \ge 0} e^{-E_n(p)\tau}  \Phi_n(p)
\end{equation}
with the $n$-th electron energy level $E_n(p)$ and the $n$-th unnormalized electron state $\Phi_n(p)$.
Because the ground state energy $E_0(p)$ is smaller than the excited state energies $E_{n>0}(p)$,  the ground state $\Phi_0(p)$ survives and the excited states $\Phi_{n>0}(p)$ die out in the infinite imaginary-time limit.
Thus the ground state is obtained by the projection
\begin{equation}
\Phi_0(p) = \lim_{\tau \to \infty} \phi (p,\tau)
\end{equation}
up to normalization.
In practical simulations, the infinite imaginary-time limit is replaced by finite large imaginary time.

The ground state defines the Berry connection
\begin{equation}
\label{eqA}
A_j (p) = -i \Phi_0^\dagger (p) \frac{\partial}{\partial p_j} \Phi_0(p)
\end{equation}
and the Berry curvature
\begin{equation}
\label{eqF}
 F(p) = \frac{\partial}{\partial p_i} A_j (p) - \frac{\partial}{\partial p_j} A_i (p)
\end{equation}
in the $(p_i,p_j)$ plane.
In practical simulations, spatial lattice size is finite and thus momentum space is also discretized.
The momentum discretization is $ \delta p = 2\pi/L$ for the spatial lattice  size $L$.
The Berry connection and curvature are formulated on this momentum lattice \cite{Fukui:2005wr,2014PhRvB..90o5443Y}.
The Berry connection is given by the link variable
\begin{equation}
U_j(p) = e^{iA_j(p)} = \frac{\Phi_0^\dagger(p) \Phi_0(p+\hat{p}_j)}{|\Phi_0^\dagger(p) \Phi_0(p+\hat{p}_j)|}
,
\end{equation}
which is an element of local $U(1)$ gauge group.
The denominator in the right-hand side is a normalization factor.
The Berry curvature is given by the plaquette
\begin{equation}
P(p) = e^{iF(p)} = U_i(p) U_j(p+\hat{p}_i) U^*_i(p+\hat{p}_j) U^*_j(p)
,
\end{equation}
which is a gauge invariant observable.
The symbol $\hat{p}_j$ denotes the unit lattice vector in the $p_j$ direction.
The Chern number is given by the integral of the Berry curvature
\begin{equation}
\label{eqN}
N = \frac{1}{2\pi} \sum_p F (p) = \frac{1}{2\pi} \sum_p {\rm Im} \ln P(p)
.
\end{equation}
The Chern number takes quantized values depending on the wave function topology.

The above formalism can be implemented to quantum Monte Carlo simulation.
The Monte Carlo configurations are generated by the standard path-integral Monte Carlo algorithm based on Eq.~\eqref{eqZprime}.
After that, the Chern number \eqref{eqN} is computed for each configuration.
Taking the ensemble average over the configurations, we obtain the quantum expectation value $\langle N \rangle$ including interaction effects.

\section{Two dimensions}
\label{sec3}

We consider the Wilson-Dirac model on the two-dimensional cubic lattice, which belongs to the class A topological insulator \cite{Schnyder:2008tya}.
The non-interacting Hamiltonian in momentum space is
\begin{equation}
\begin{split}
  H_0(p) =\ & t \sigma_1 \sin p_1 + t \sigma_2 \sin p_2
\\
& + \sigma_3 ( t \cos p_1 + t \cos p_2 + m)
\end{split}
\end{equation}
with the hopping parameter $t$ and the mass parameter $m$.
The corresponding Dirac operator is
\begin{equation}
\label{eqK2D}
\begin{split}
&K(x,\tau|x',\tau)
\\
=& \left\{ \frac{\partial}{\partial \tau}  + i\eta(x,\tau) - m\sigma_3 \right\} \delta_{x,x'}
\\
&- \frac{t}{2} \sum_{j=1,2} \bigg\{ \left( \sigma_3-i\sigma_j \right) \delta_{x+\hat{j},x'}
 + \left( \sigma_3+i\sigma_j \right) \delta_{x-\hat{j},x'} \bigg\}
.
\end{split}
\end{equation}
The symbol $\hat{j}$ denotes the unit lattice vector in the $x_j$ direction.
Since a chemical potential is not introduced, the Dirac operator has particle-hole symmetry, and thus the system is half filling.

In practical simulations, imaginary time $\tau$ is discretized with a small discretization parameter $\delta \tau$.
The discretized action is
\begin{equation}
\begin{split}
S'
=& \ \delta\tau \sum_{\tau,\tau'} \sum_{x,x'} \Bigg[ \sum_{s=\uparrow,\downarrow} \psi^\dagger_s (x,\tau) K(x,\tau|x',\tau') \psi_s (x',\tau') 
\\
&+ \frac{1}{2} V^{-1}(x|x') \eta(x,\tau)  \eta(x',\tau) \delta_{\tau,\tau'} \Bigg]
\end{split}
\end{equation}
with the discretized Dirac operator
\begin{equation}
\label{eqK2Ddt}
\begin{split}
&K(x,\tau|x',\tau')
\\
=& \left( \frac{1}{\delta\tau} - m \right) \sigma_3 \delta_{\tau,\tau'} \delta_{x,x'}
\\
&- \frac{1}{2\delta\tau} \bigg\{ \left( \sigma_3-1 \right) e^{i \delta \tau \eta(x,\tau)} \delta_{\tau+\delta \tau,\tau'}
\\
&+ \left( \sigma_3+1 \right) e^{-i \delta \tau \eta(x',\tau')} \delta_{\tau-\delta\tau,\tau'} \bigg\} \delta_{x,x'}
\\
&- \frac{t}{2} \sum_{j=1,2} \bigg\{ \left( \sigma_3-i\sigma_j \right) \delta_{x+\hat{j},x'}
+ \left( \sigma_3+i\sigma_j \right) \delta_{x-\hat{j},x'} \bigg\} \delta_{\tau,\tau'}
.
\end{split}
\end{equation}
Here the imaginary-time derivative is discretized by the Wilson fermion formalism to avoid unphysical doublers \cite{Rothe:1992nt, *Montvay:1994cy}.
The discretized imaginary-time derivative is similar to the spatial hopping terms, but their coefficients are different.
In the non-relativistic Wilson-Dirac model, the coefficients must satisfy $\delta\tau \ll 1/t$ to avoid the unphysical doublers in the imaginary-time direction.
At $t = 1/\delta\tau$, this action corresponds to the relativistic Wilson fermion action.
Since the auxiliary field is like the imaginary-time component of the gauge potential in Eq.~\eqref{eqK2D}, it is implemented as the imaginary-time component of the link variable in Eq.~\eqref{eqK2Ddt}.

\begin{figure}[t]
 \includegraphics[width=.49\textwidth]{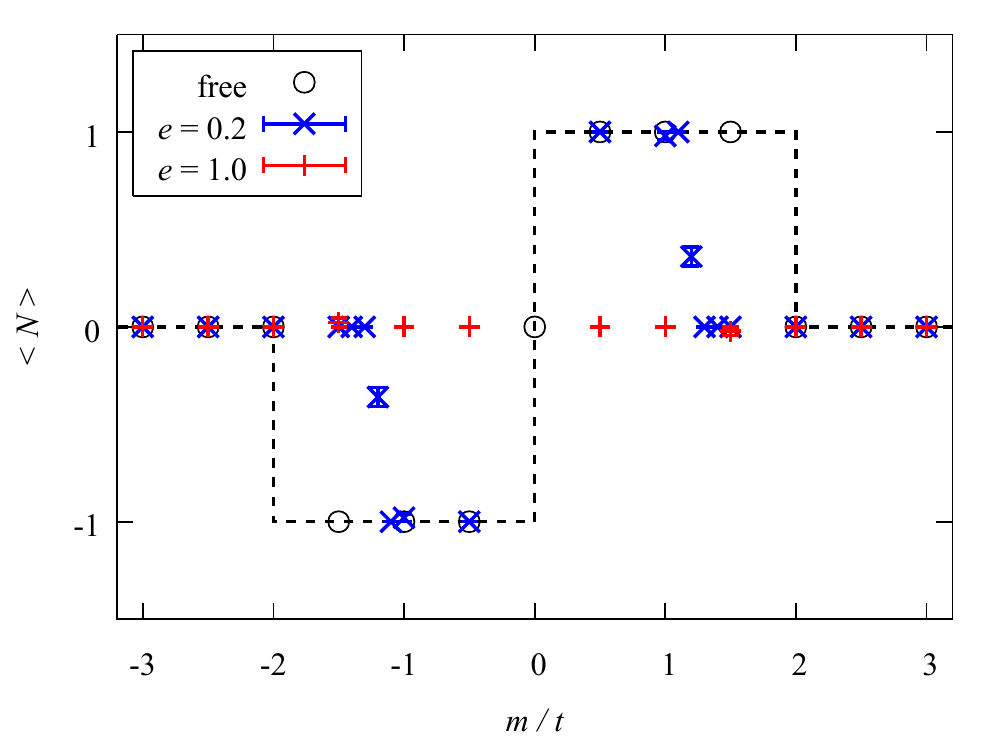}
 \caption{\label{figN2D} 
The Chern number $\langle N\rangle$.
The black circles are the numerical solutions and the black dashed line is the analytical solution of the non-interacting case.
The colored symbols are quantum Monte Carlo simulation results.
}
\end{figure}

\begin{figure*}[t]
\begin{tabular}{c}
 \includegraphics[width=.48\textwidth]{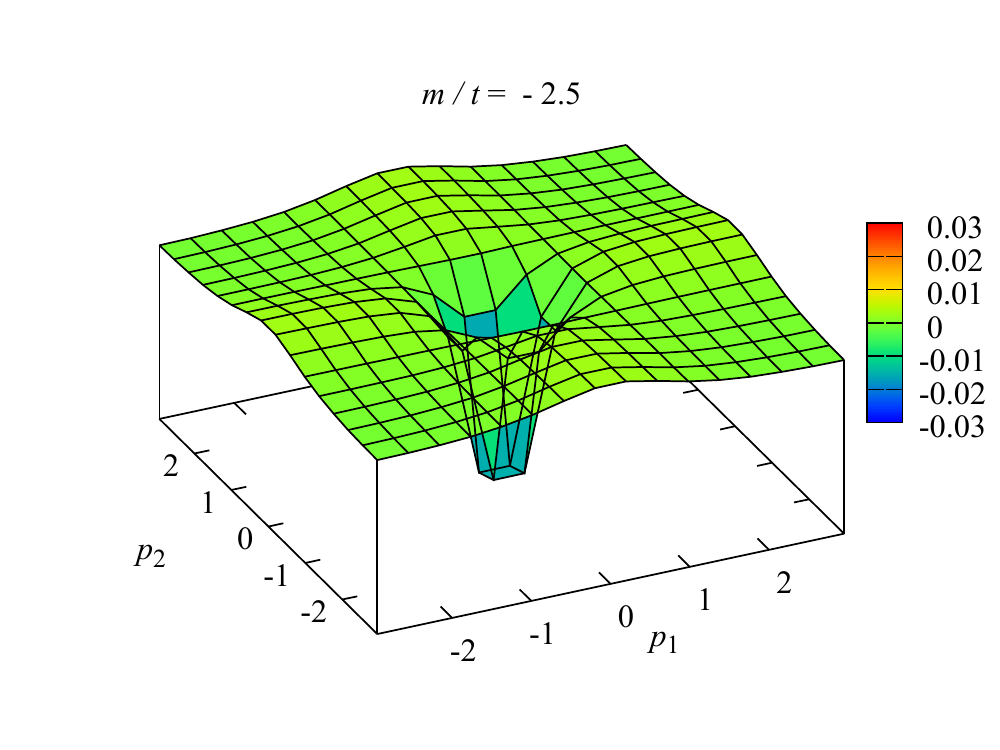}
 \includegraphics[width=.48\textwidth]{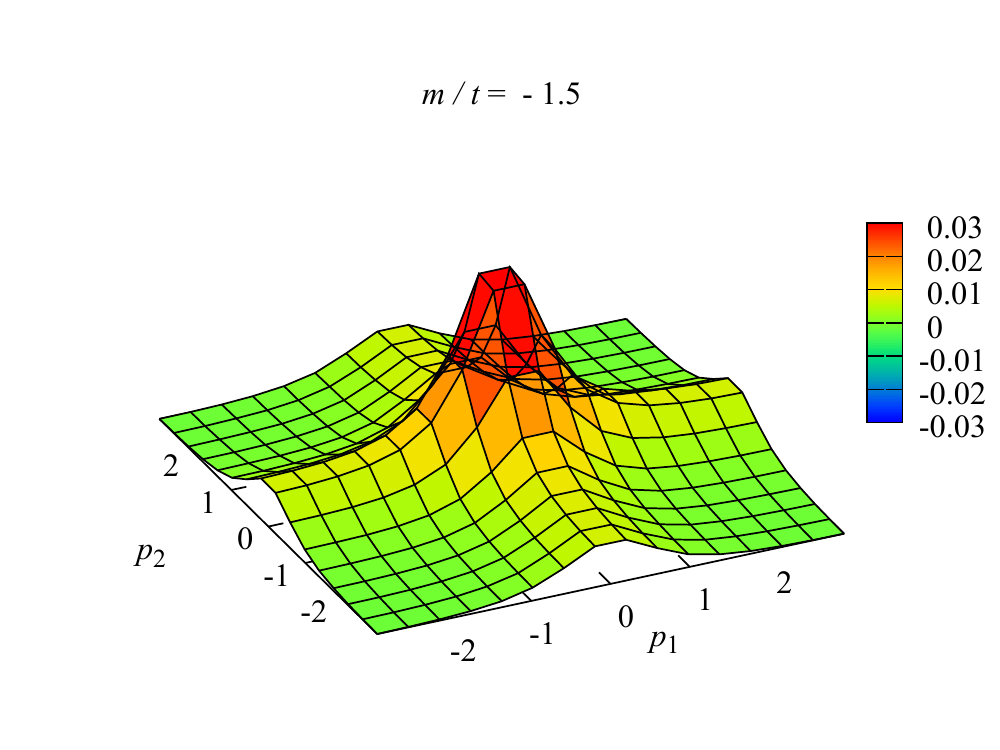}
\\
 \includegraphics[width=.48\textwidth]{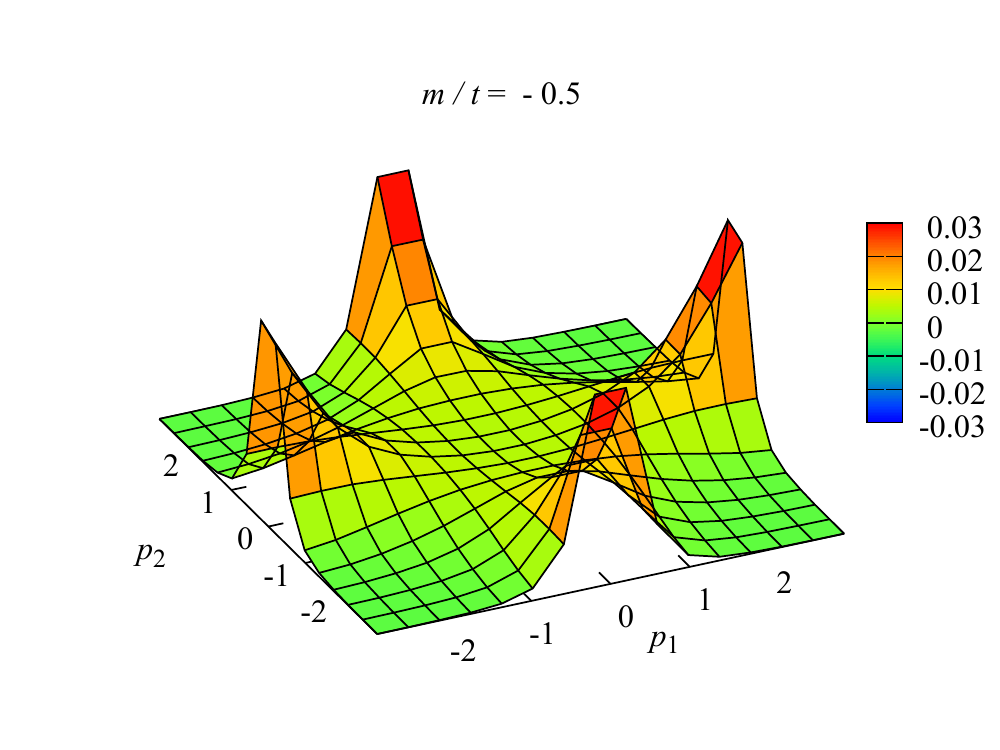}
 \includegraphics[width=.48\textwidth]{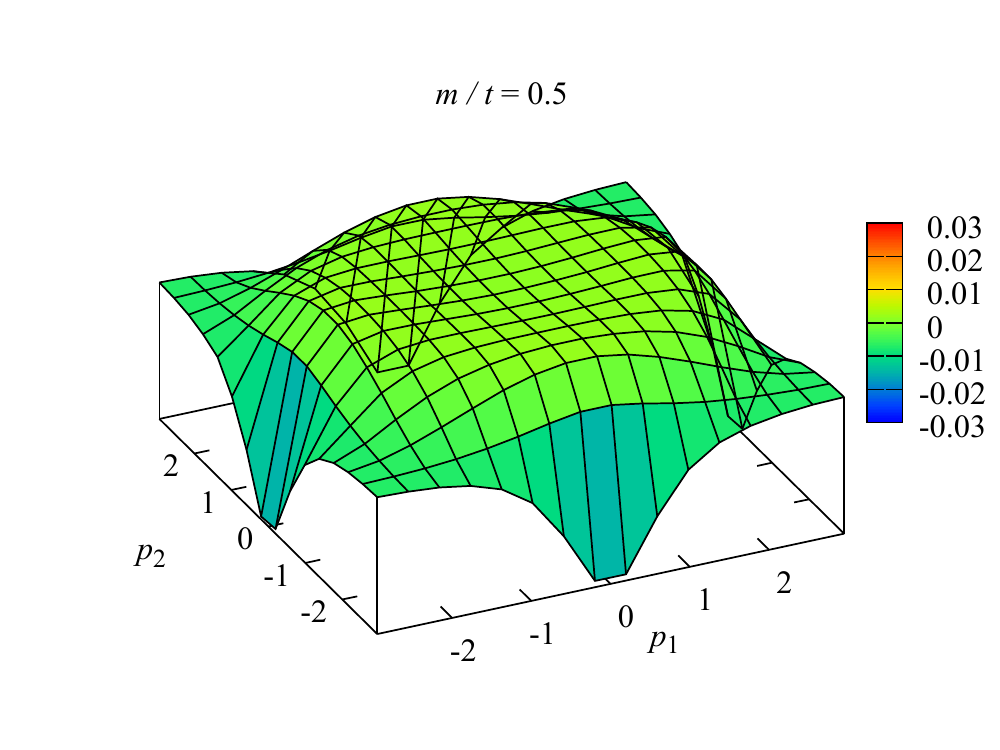}
\\
 \includegraphics[width=.48\textwidth]{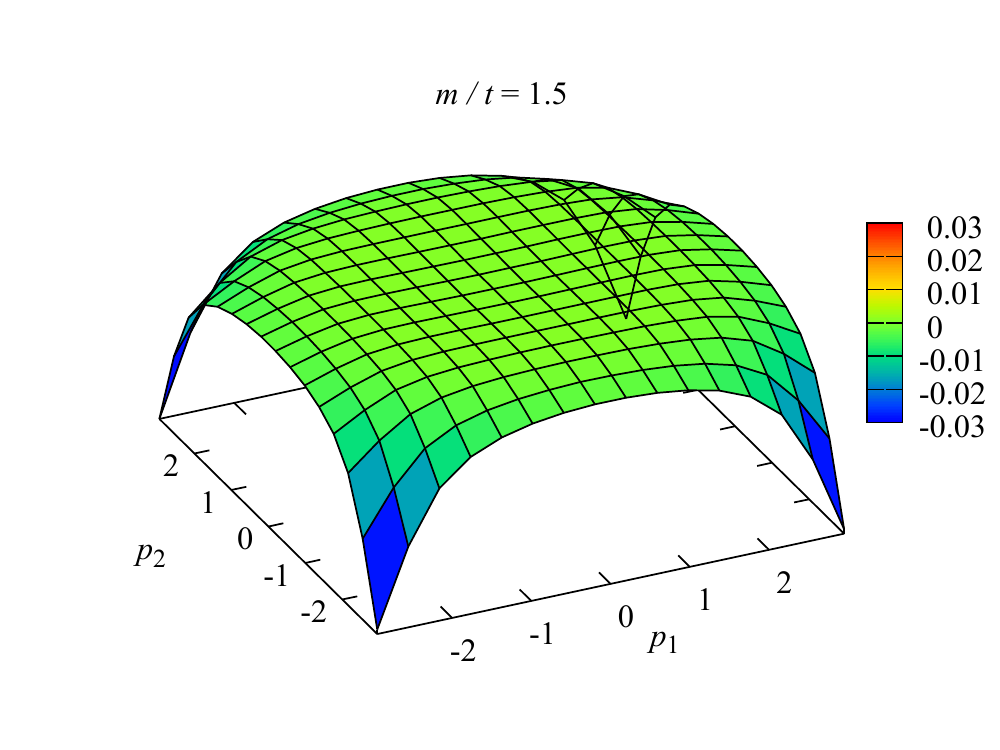}
 \includegraphics[width=.48\textwidth]{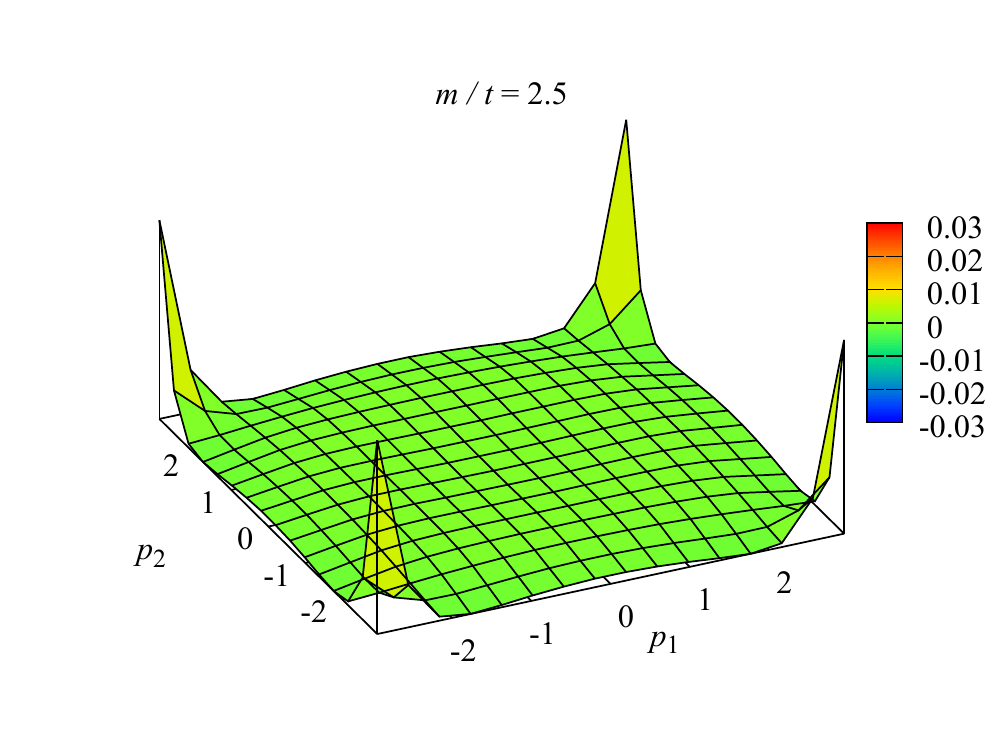}
\end{tabular}
 \caption{\label{figA} 
The Berry curvature $F(p)$ in the non-interacting case.
The momentum space $(p_1,p_2)$ is periodic.
 }
\end{figure*}

Before the quantum Monte Carlo simulation of the interacting case, we study the non-interacting case.
The Dirac operator $K$ in Eq.~\eqref{eqphipt} is replaced by the non-interacting Dirac operator $K_0$.
The numerical solution is unique and the Monte Carlo sampling is not necessary.
The spatial lattice size is $L_x L_y = 16^2$ and the number of temporal discretization is $L_\tau = 50$.
The boundary conditions are periodic in spatial boundaries and anti-periodic in the imaginary-time boundary.
The hopping parameter is fixed at $t=0.2/\delta\tau$.
We have numerically checked the parameter independence of the following results.

The Wilson-Dirac model has quantum phase transitions as the mass parameter changes.
In the non-interacting case, the mass dependence of the Chern number is analytically calculated \cite{Jansen:1992tw,Golterman:1992ub,Qi:2008ew}.
The results of the numerical simulation and the analytical calculation are shown in Fig.~\ref{figN2D}.
The numerical simulation completely reproduces the analytical calculation.
There are topological insulator phase with $N=\pm 1$ in $|m/t| <2$ and normal insulator phase with $N=0$ in $|m/t| \ge 2$.
The distribution of the Berry curvature is shown in Fig.~\ref{figA}.
Comparing Fig.~\ref{figN2D} and Fig.~\ref{figA}, we clearly see the relation between the change of the Chern number and  the peak structure of the Berry curvature.
The Berry curvature at $m/t = 2.5$ has a negative peak at $(p_1,p_2) = (0,0)$ and the Berry curvature at $m/t = 1.5$ has a positive peak at $(p_1,p_2) = (0,0)$.
This change relates to the gapless mode at $(p_1,p_2) = (0,0)$ and the topological transition at $m/t = 2$.
In the same way, the peaks at $(p_1,p_2) = (0,\pi)$ and $(\pi,0)$ relates to the transition at $m/t = 0$, and the peak at $(p_1,p_2) = (\pi,\pi)$ relates to the transition at $m/t = -2$.

Then we introduce the electron-electron interaction by the quantum Monte Carlo method.
We adopted the long-range Coulomb interaction
\begin{equation}
\label{eqCoulomb}
 V(x|x') = \frac{e^2}{\sqrt{\sum_j(x_j-x'_j)^2}+\epsilon}
\end{equation}
with the short-range cutoff $\epsilon=0.1 \delta\tau$ to avoid the on-site singularity at $x=x'$.
We adopted the quenched approximation, where the quantum effects by fermion loops are neglected, to reduce simulation cost \cite{[{See textbooks: }]Rothe:1992nt, *Montvay:1994cy}.
The quenched approximation is a familiar scheme for the semi-quantitative study in the path-integral Monte Carlo simulation.
Even though the Dirac operator is not positive definite, there is no sign problem in the quenched approximation.
The simulation parameters are the same as in the non-interacting case.

\begin{figure}[t]
 \includegraphics[width=.49\textwidth]{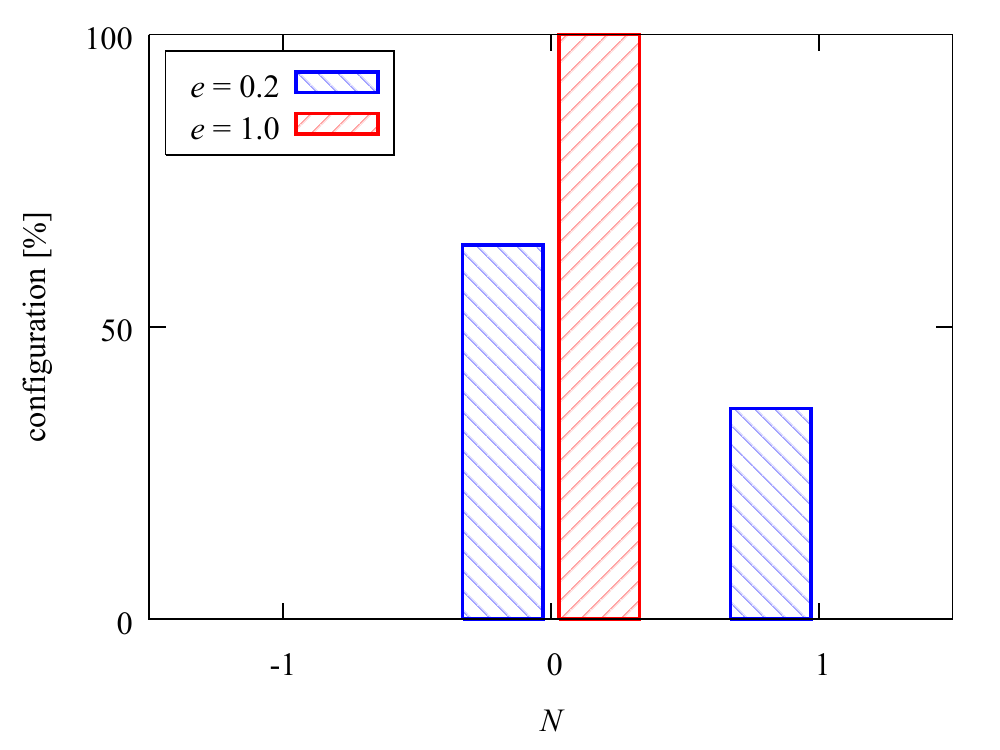}
 \caption{\label{figH} 
The configuration numbers of the Chern number $N$ in the Monte Carlo ensemble with $m/t = 1.2$.
 }
\end{figure}

The Chern numbers in the interacting cases are shown in Fig.~\ref{figN2D}.
In the weakly interacting case $e = 0.2$, the phase transition points shift from $|m_c/t| = 2.0$ to $|m_c/t| \sim 1.2$, and thus the topological insulator phase shrinks.
This means that the electron-electron interaction tends to destroy non-trivial topology.
We remark that the critical point at $m_c/t=0$ is not affected by the interaction because it is protected by the inversion symmetry.
In the strongly interacting case $e = 1.0$, non-trivial topology is completely destroyed and only the topologically trivial phase is observed.
In this case, we have nonzero condensate $\left<\bar{\psi}\psi\right> := \left<\psi^\dag \sigma_3 \psi\right> \neq 0$, which is consistent with the the charge-density wave transition in the interacting Haldane model~\cite{Varney:2010PRB,*Varney:2011PRB}.
Another interesting point is the non-integer value of the ensemble average near the phase transition.
In Fig.~\ref{figN2D}, the data of $e = 0.2$ show $\langle N\rangle \sim 0.4$ at $m/t=1.2$.
While the Chern number $N$ in each configuration is an integer, the ensemble average $\langle N\rangle$ is not.
The histogram of the configurations is shown in Fig.~\ref{figH}.
The configurations with different Chern numbers are strongly mixed near the phase transition.
This means that topology is quite unstable due to quantum fluctuation.
The calculation was done in a finite volume $L_xL_y=16$.
Whether this fluctuation survives in the thermodynamics limit depends on the order of phase transition.
It will be clarified by the finite-size scaling analysis of topological susceptibility \cite{SMIT198836,VINK1988549}

\section{Three dimensions}
\label{sec4}

\begin{figure}[t]
 \includegraphics[width=.3\textwidth]{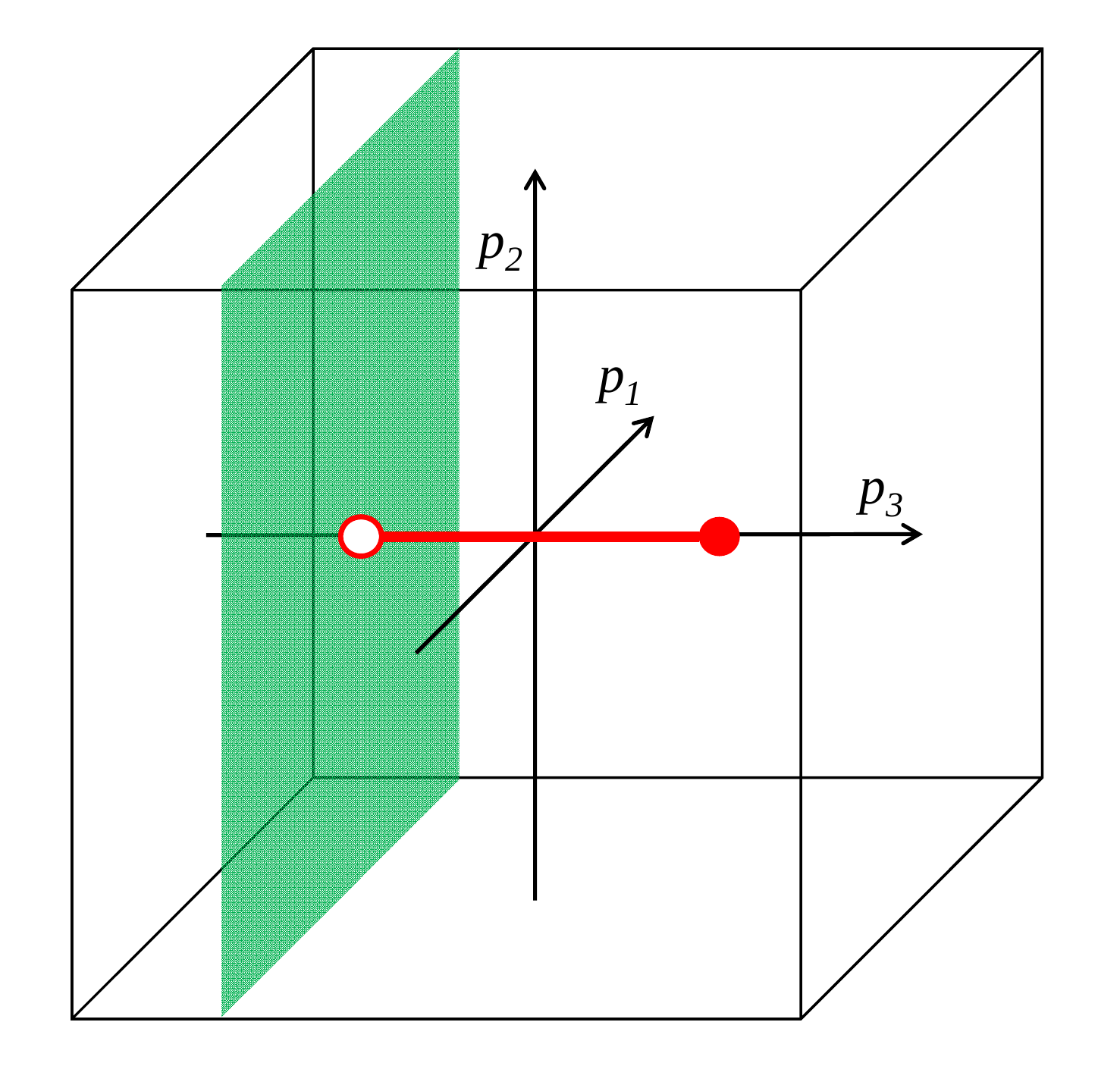}
 \caption{\label{figWeyl} 
Schematic figure of Weyl points (red circles).
The Chern number is calculated in the $(p_1,p_2)$ plane with $p_3$ fixed (green sheet).
 }
\end{figure}

The above analysis can be applied to Weyl semimetals in three dimensions.
We consider the non-interacting Hamiltonian
\begin{equation}
\begin{split}
 H_0(p) =
&\  t \sigma_1 \sin p_1 + t \sigma_2 \sin p_2 
\\
&+ t \sigma_3 ( \cos p_1 + \cos p_2 + \cos p_3 - 2)
.
\end{split}
\end{equation}
The structure of momentum space is schematically drawn in Fig.~\ref{figWeyl}.
There are two Weyl points at $(p_1,p_2,p_3) = (0,0,\pm \pi/2)$, which play roles of a monopole and an anti-monopole.
We calculate the Chern number in the $(p_1,p_2)$ plane with $p_3$ fixed.
The Chern number is equal to the number of topological fluxes penetrating the plane.
Since the flux exists only between the two Weyl points, we can search the positions of the Weyl points by calculating the Chern number as a function of $p_3$.

Theoretical formulation is the same as the two-dimensional case, except for the existence of the $x_3$ direction.
The action is
\begin{equation}
\begin{split}
S'
=& \int d\tau \sum_{x,x'} \Bigg[ \sum_{s=\uparrow,\downarrow} \psi^\dagger_s (x,\tau) K(x,\tau|x',\tau) \psi_s(x',\tau)
\\
&+ \frac{1}{2} V^{-1}(x|x') \eta(x,\tau)  \eta(x',\tau) \Bigg]
\end{split}
\end{equation}
with
\begin{equation}
\begin{split}
&K(x,\tau|x',\tau)
\\
=& \left\{ \frac{\partial}{\partial \tau}  + i\eta(x,\tau) + 2t \sigma_3 \right\} \delta_{x,x'}
\\
&- \frac{t}{2} \sum_{j=1,2} \bigg\{ \left( \sigma_3-i\sigma_j \right) \delta_{x+\hat{j},x'}
 + \left( \sigma_3+i\sigma_j \right) \delta_{x-\hat{j},x'} \bigg\}
\\
&- \frac{t}{2} \sigma_3 \bigg( \delta_{x+\hat{3},x'} + \delta_{x-\hat{3},x'} \bigg)
.
\end{split}
\end{equation}
The auxiliary field is now in three spatial dimensions.
Simulation scheme is also the same.
The spatial lattice volume is $L_x L_y \times L_z = 8^2 \times 16$, and the temporal size is $L_\tau = 50$ as before.
Other conditions are the same as the two-dimensional case.

\begin{figure}[t]
 \includegraphics[width=.49\textwidth]{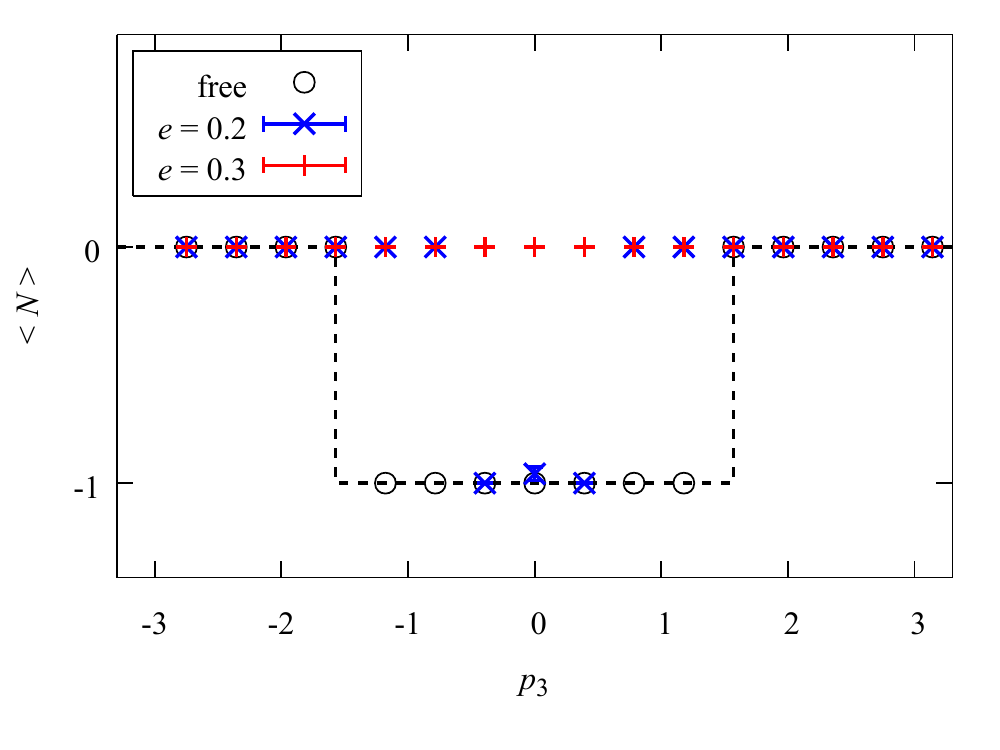}
 \caption{\label{figN3D} 
The Chern number $\langle N\rangle$.
The black circles are the numerical solutions and the dashed line is the analytical solution of the non-interacting case.
The colored symbols are the quantum Monte Carlo simulation results.
 }
\end{figure}

\begin{figure}[t]
 \includegraphics[width=.49\textwidth]{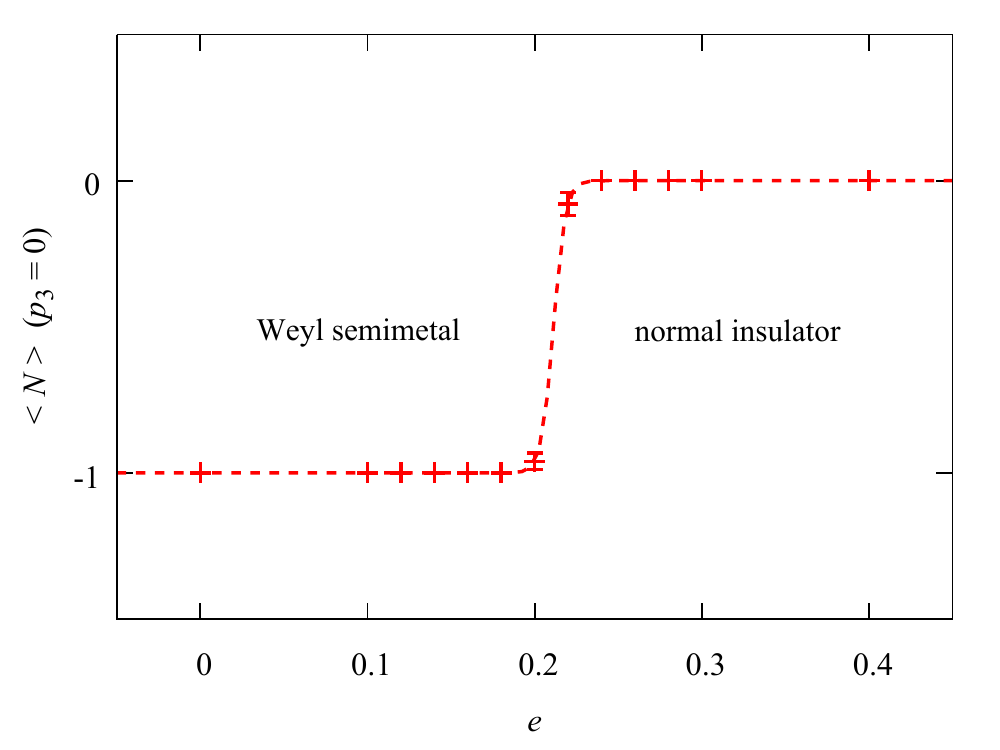}
 \caption{\label{figE} 
Phase diagram.
The Chern number $\langle N\rangle$ at $p_3=0$ is plotted as a topological order parameter.
The dashed curve is a hyperbolic tangent function with $e_c \sim 0.21$.}
\end{figure}

The Chern number in the $(p_1,p_2)$ plane as a function of $p_3$ is shown in Fig.~\ref{figN3D}.
The point where the Chern number changes is identified as a Weyl point.
In the non-interacting case, the Weyl points are located at $p_3 = \pm \pi/2$.
In the interacting case with $e=0.2$, the Weyl points shift to $p_3 \sim \pm \pi/4$, and with $e=0.3$, the Weyl points disappear, which is consistent with the mean-field analysis~\cite{Sekine:2013pda,Maciejko:2013lua}.
Such a shift and breakdown of Weyl points was also discussed with the short-range interaction~\cite{Dora:2013PRB,Witczak-Krempa:2014nva}.
In general, when a pair of two Weyl points disappears, there are two possibilities: the formation of one Dirac point or the transition to a normal insulator.
We calculated the electron energy levels $E_n(p)$ from the exponents of the electron propagator and found nonzero energy gap at $p=0$.
Thus we conclude that this is a semimetal-insulator transition.
Regarding the Chern number at $p_3=0$ as the order parameter of this transition, we draw the phase diagram of this model in Fig.~\ref{figE}.
In the strongly coupling region of $e > e_c \sim 0.21$, the interaction washes out the Weyl points and induces a normal insulator.

\section{Summary}
\label{sec5}

We studied topological phase transitions by the path-integral Monte Carlo simulation.
Firstly, we analyzed the Wilson-Dirac model in two dimensions.
The phase transition point is shifted or vanished by the electron-electron interaction.
The quantum fluctuation of the Chern number is observed near the phase transition.
Secondly, we analyzed a model of Weyl semimetals in three dimensions.
Weyl points are shifted or smeared out by the electron-electron interaction.
Although these analyses were done in simple models, the same analysis is possible for more realistic situations.
The applications to other lattice structures, other symmetry classes, and other interactions, are straightforward.
It would be also interesting to study the edge/surface state using our framework.

Our result predicts that the bulk dispersion is modified due to the interaction effect.
Although it is in general difficult to see the bulk spectrum directly, the surface state spectrum can be experimentally observed.
In particular for Weyl semimetals, one can detect the Fermi arc and its end point is identified as the bulk Weyl point.
In this way it would be possible to observe the shift of bulk spectrum in experiments.

We finally remark that the shift behavior of the topological critical point, shown in this work, has a close connection with the Aoki phase in the lattice QCD~\cite{Aoki:1983qi,*Aoki:1985mk,*Aoki:1986kt,*Aoki:1986xr,*Aoki:1987us}, which appears in the interacting lattice fermion system~\cite{[{It was pointed out in the context of topological phases by }] Araki:2013dsa}.
In the Aoki phase, the (flavor-)parity symmetry is spontaneously broken, and the topological number cannot be defined since it is not gapped anymore.
Thus the Aoki phase structure plays an important role in practice to construct a lattice chiral fermion as a topological edge state~\cite{[{For the connection with topological phases, see }] [{ and references therein.}] Kimura:2015ixh}.
One possible way to reduce the Aoki phase region, in other words, to spread the topologically non-trivial region, is the twisted mass formalism~\cite{[{See a review article: }] Frezzotti:2002iv}.
In the context of condensed-matter physics, it could be realized as a spin-dependent mass term, induced by the spin-orbit interaction.
This implies that the shift behavior of the topological critical point due to the electron interaction would be suppressed by such a spin-dependent interaction.

\begin{acknowledgments}
AY was supported by JSPS KAKENHI Grant No.~JP15K17624.   
TK was supported by the MEXT-Supported Program for the Strategic Research Foundation at Private Universities ``Topological Science'' (No.~S1511006), and JSPS Grant-in-Aid for Scientific Research on Innovative Areas ``Topological Materials Science'' (No.~JP15H05855).
The numerical simulations were carried out on SX-ACE in Osaka University.
\end{acknowledgments}

\bibliographystyle{apsrev4-1}
\bibliography{top_ph}

\begin{thebibliography}{45}%
\makeatletter
\providecommand \@ifxundefined [1]{%
 \@ifx{#1\undefined}
}%
\providecommand \@ifnum [1]{%
 \ifnum #1\expandafter \@firstoftwo
 \else \expandafter \@secondoftwo
 \fi
}%
\providecommand \@ifx [1]{%
 \ifx #1\expandafter \@firstoftwo
 \else \expandafter \@secondoftwo
 \fi
}%
\providecommand \natexlab [1]{#1}%
\providecommand \enquote  [1]{``#1''}%
\providecommand \bibnamefont  [1]{#1}%
\providecommand \bibfnamefont [1]{#1}%
\providecommand \citenamefont [1]{#1}%
\providecommand \href@noop [0]{\@secondoftwo}%
\providecommand \href [0]{\begingroup \@sanitize@url \@href}%
\providecommand \@href[1]{\@@startlink{#1}\@@href}%
\providecommand \@@href[1]{\endgroup#1\@@endlink}%
\providecommand \@sanitize@url [0]{\catcode `\\12\catcode `\$12\catcode
  `\&12\catcode `\#12\catcode `\^12\catcode `\_12\catcode `\%12\relax}%
\providecommand \@@startlink[1]{}%
\providecommand \@@endlink[0]{}%
\providecommand \url  [0]{\begingroup\@sanitize@url \@url }%
\providecommand \@url [1]{\endgroup\@href {#1}{\urlprefix }}%
\providecommand \urlprefix  [0]{URL }%
\providecommand \Eprint [0]{\href }%
\providecommand \doibase [0]{http://dx.doi.org/}%
\providecommand \selectlanguage [0]{\@gobble}%
\providecommand \bibinfo  [0]{\@secondoftwo}%
\providecommand \bibfield  [0]{\@secondoftwo}%
\providecommand \translation [1]{[#1]}%
\providecommand \BibitemOpen [0]{}%
\providecommand \bibitemStop [0]{}%
\providecommand \bibitemNoStop [0]{.\EOS\space}%
\providecommand \EOS [0]{\spacefactor3000\relax}%
\providecommand \BibitemShut  [1]{\csname bibitem#1\endcsname}%
\let\auto@bib@innerbib\@empty
\bibitem [{\citenamefont {Schnyder}\ \emph {et~al.}(2008)\citenamefont
  {Schnyder}, \citenamefont {Ryu}, \citenamefont {Furusaki},\ and\
  \citenamefont {Ludwig}}]{Schnyder:2008tya}%
  \BibitemOpen
  \bibfield  {author} {\bibinfo {author} {\bibfnamefont {A.~P.}\ \bibnamefont
  {Schnyder}}, \bibinfo {author} {\bibfnamefont {S.}~\bibnamefont {Ryu}},
  \bibinfo {author} {\bibfnamefont {A.}~\bibnamefont {Furusaki}}, \ and\
  \bibinfo {author} {\bibfnamefont {A.~W.~W.}\ \bibnamefont {Ludwig}},\ }\href
  {\doibase 10.1103/PhysRevB.78.195125} {\bibfield  {journal} {\bibinfo
  {journal} {Phys. Rev.}\ }\textbf {\bibinfo {volume} {B78}},\ \bibinfo {pages}
  {195125} (\bibinfo {year} {2008})},\ \Eprint {http://arxiv.org/abs/0803.2786}
  {arXiv:0803.2786 [cond-mat.mes-hall]} \BibitemShut {NoStop}%
\bibitem [{\citenamefont {Kitaev}(2009)}]{Kitaev:2009mg}%
  \BibitemOpen
  \bibfield  {author} {\bibinfo {author} {\bibfnamefont {A.~Y.}\ \bibnamefont
  {Kitaev}},\ }\href {\doibase 10.1063/1.3149495} {\bibfield  {journal}
  {\bibinfo  {journal} {AIP Conf. Proc.}\ }\textbf {\bibinfo {volume} {1134}},\
  \bibinfo {pages} {22} (\bibinfo {year} {2009})},\ \Eprint
  {http://arxiv.org/abs/0901.2686} {arXiv:0901.2686 [cond-mat.mes-hall]}
  \BibitemShut {NoStop}%
\bibitem [{\citenamefont {Fidkowski}\ and\ \citenamefont
  {Kitaev}(2010)}]{Fidkowski:2009dba}%
  \BibitemOpen
  \bibfield  {author} {\bibinfo {author} {\bibfnamefont {L.}~\bibnamefont
  {Fidkowski}}\ and\ \bibinfo {author} {\bibfnamefont {A.}~\bibnamefont
  {Kitaev}},\ }\href {\doibase 10.1103/PhysRevB.81.134509} {\bibfield
  {journal} {\bibinfo  {journal} {Phys. Rev.}\ }\textbf {\bibinfo {volume}
  {B81}},\ \bibinfo {pages} {134509} (\bibinfo {year} {2010})},\ \Eprint
  {http://arxiv.org/abs/0904.2197} {arXiv:0904.2197 [cond-mat.str-el]}
  \BibitemShut {NoStop}%
\bibitem [{\citenamefont {Pesin}\ and\ \citenamefont
  {Balents}(2010)}]{Pesin:2009NP}%
  \BibitemOpen
  \bibfield  {author} {\bibinfo {author} {\bibfnamefont {D.~A.}\ \bibnamefont
  {Pesin}}\ and\ \bibinfo {author} {\bibfnamefont {L.}~\bibnamefont
  {Balents}},\ }\href {\doibase 10.1038/nphys1606} {\bibfield  {journal}
  {\bibinfo  {journal} {Nat. Phys.}\ }\textbf {\bibinfo {volume} {6}},\
  \bibinfo {pages} {376} (\bibinfo {year} {2010})},\ \Eprint
  {http://arxiv.org/abs/0907.2962} {arXiv:0907.2962 [cond-mat.str-el]}
  \BibitemShut {NoStop}%
\bibitem [{\citenamefont {Takimoto}(2011)}]{Takimoto:2011JPSJ}%
  \BibitemOpen
  \bibfield  {author} {\bibinfo {author} {\bibfnamefont {T.}~\bibnamefont
  {Takimoto}},\ }\href {\doibase 10.1143/JPSJ.80.123710} {\bibfield  {journal}
  {\bibinfo  {journal} {J. Phys. Soc. Jap.}\ }\textbf {\bibinfo {volume}
  {80}},\ \bibinfo {pages} {123710} (\bibinfo {year} {2011})}\BibitemShut
  {NoStop}%
\bibitem [{\citenamefont {Dzero}\ \emph {et~al.}(2012)\citenamefont {Dzero},
  \citenamefont {Sun}, \citenamefont {Coleman},\ and\ \citenamefont
  {Galitski}}]{Dzero:2012PRB}%
  \BibitemOpen
  \bibfield  {author} {\bibinfo {author} {\bibfnamefont {M.}~\bibnamefont
  {Dzero}}, \bibinfo {author} {\bibfnamefont {K.}~\bibnamefont {Sun}}, \bibinfo
  {author} {\bibfnamefont {P.}~\bibnamefont {Coleman}}, \ and\ \bibinfo
  {author} {\bibfnamefont {V.}~\bibnamefont {Galitski}},\ }\href {\doibase
  10.1103/PhysRevB.85.045130} {\bibfield  {journal} {\bibinfo  {journal} {Phys.
  Rev.}\ }\textbf {\bibinfo {volume} {B85}},\ \bibinfo {pages} {045130}
  (\bibinfo {year} {2012})},\ \Eprint {http://arxiv.org/abs/1108.3371}
  {arXiv:1108.3371 [cond-mat.str-el]} \BibitemShut {NoStop}%
\bibitem [{\citenamefont {Sheng}\ \emph {et~al.}(2011)\citenamefont {Sheng},
  \citenamefont {Gu}, \citenamefont {Sun},\ and\ \citenamefont
  {Sheng}}]{Sheng:2011NC}%
  \BibitemOpen
  \bibfield  {author} {\bibinfo {author} {\bibfnamefont {D.}~\bibnamefont
  {Sheng}}, \bibinfo {author} {\bibfnamefont {Z.-C.}\ \bibnamefont {Gu}},
  \bibinfo {author} {\bibfnamefont {K.}~\bibnamefont {Sun}}, \ and\ \bibinfo
  {author} {\bibfnamefont {L.}~\bibnamefont {Sheng}},\ }\href {\doibase
  10.1038/ncomms1380} {\bibfield  {journal} {\bibinfo  {journal} {Nature
  Comm.}\ }\textbf {\bibinfo {volume} {2}},\ \bibinfo {pages} {389} (\bibinfo
  {year} {2011})},\ \Eprint {http://arxiv.org/abs/1102.2658} {arXiv:1102.2658
  [cond-mat.str-el]} \BibitemShut {NoStop}%
\bibitem [{\citenamefont {Venderbos}\ \emph {et~al.}(2012)\citenamefont
  {Venderbos}, \citenamefont {Kourtis}, \citenamefont {van~den Brink},\ and\
  \citenamefont {Daghofer}}]{Venderbos:2012PRL}%
  \BibitemOpen
  \bibfield  {author} {\bibinfo {author} {\bibfnamefont {J.~W.~F.}\
  \bibnamefont {Venderbos}}, \bibinfo {author} {\bibfnamefont {S.}~\bibnamefont
  {Kourtis}}, \bibinfo {author} {\bibfnamefont {J.}~\bibnamefont {van~den
  Brink}}, \ and\ \bibinfo {author} {\bibfnamefont {M.}~\bibnamefont
  {Daghofer}},\ }\href {\doibase 10.1103/PhysRevLett.108.126405} {\bibfield
  {journal} {\bibinfo  {journal} {Phys. Rev. Lett.}\ }\textbf {\bibinfo
  {volume} {108}},\ \bibinfo {pages} {126405} (\bibinfo {year} {2012})},\
  \Eprint {http://arxiv.org/abs/1109.5955} {arXiv:1109.5955 [cond-mat.str-el]}
  \BibitemShut {NoStop}%
\bibitem [{\citenamefont {Hohenadler}\ and\ \citenamefont
  {Assaad}(2013)}]{Hohenadler:2013JPC}%
  \BibitemOpen
  \bibfield  {author} {\bibinfo {author} {\bibfnamefont {M.}~\bibnamefont
  {Hohenadler}}\ and\ \bibinfo {author} {\bibfnamefont {F.~F.}\ \bibnamefont
  {Assaad}},\ }\href {\doibase 10.1088/0953-8984/25/14/143201} {\bibfield
  {journal} {\bibinfo  {journal} {J. Phys.: Cond. Matt.}\ }\textbf {\bibinfo
  {volume} {25}},\ \bibinfo {pages} {143201} (\bibinfo {year} {2013})},\
  \Eprint {http://arxiv.org/abs/1211.1774} {arXiv:1211.1774 [cond-mat.str-el]}
  \BibitemShut {NoStop}%
\bibitem [{\citenamefont {{Wehling}}\ \emph {et~al.}(2011)\citenamefont
  {{Wehling}}, \citenamefont {{{\c S}a{\c s}{\i}o{\v g}lu}}, \citenamefont
  {{Friedrich}}, \citenamefont {{Lichtenstein}}, \citenamefont {{Katsnelson}},\
  and\ \citenamefont {{Bl{\"u}gel}}}]{2011PhRvL.106w6805W}%
  \BibitemOpen
  \bibfield  {author} {\bibinfo {author} {\bibfnamefont {T.~O.}\ \bibnamefont
  {{Wehling}}}, \bibinfo {author} {\bibfnamefont {E.}~\bibnamefont {{{\c S}a{\c
  s}{\i}o{\v g}lu}}}, \bibinfo {author} {\bibfnamefont {C.}~\bibnamefont
  {{Friedrich}}}, \bibinfo {author} {\bibfnamefont {A.~I.}\ \bibnamefont
  {{Lichtenstein}}}, \bibinfo {author} {\bibfnamefont {M.~I.}\ \bibnamefont
  {{Katsnelson}}}, \ and\ \bibinfo {author} {\bibfnamefont {S.}~\bibnamefont
  {{Bl{\"u}gel}}},\ }\href {\doibase 10.1103/PhysRevLett.106.236805} {\bibfield
   {journal} {\bibinfo  {journal} {Phys. Rev. Lett.}\ }\textbf {\bibinfo
  {volume} {106}},\ \bibinfo {eid} {236805} (\bibinfo {year} {2011})},\ \Eprint
  {http://arxiv.org/abs/1101.4007} {arXiv:1101.4007 [cond-mat.mes-hall]}
  \BibitemShut {NoStop}%
\bibitem [{\citenamefont {Ulybyshev}\ \emph {et~al.}(2013)\citenamefont
  {Ulybyshev}, \citenamefont {Buividovich}, \citenamefont {Katsnelson},\ and\
  \citenamefont {Polikarpov}}]{Ulybyshev:2013swa}%
  \BibitemOpen
  \bibfield  {author} {\bibinfo {author} {\bibfnamefont {M.~V.}\ \bibnamefont
  {Ulybyshev}}, \bibinfo {author} {\bibfnamefont {P.~V.}\ \bibnamefont
  {Buividovich}}, \bibinfo {author} {\bibfnamefont {M.~I.}\ \bibnamefont
  {Katsnelson}}, \ and\ \bibinfo {author} {\bibfnamefont {M.~I.}\ \bibnamefont
  {Polikarpov}},\ }\href {\doibase 10.1103/PhysRevLett.111.056801} {\bibfield
  {journal} {\bibinfo  {journal} {Phys. Rev. Lett.}\ }\textbf {\bibinfo
  {volume} {111}},\ \bibinfo {pages} {056801} (\bibinfo {year} {2013})},\
  \Eprint {http://arxiv.org/abs/1304.3660} {arXiv:1304.3660 [cond-mat.str-el]}
  \BibitemShut {NoStop}%
\bibitem [{\citenamefont {Smith}\ and\ \citenamefont {von
  Smekal}(2014)}]{Smith:2014tha}%
  \BibitemOpen
  \bibfield  {author} {\bibinfo {author} {\bibfnamefont {D.}~\bibnamefont
  {Smith}}\ and\ \bibinfo {author} {\bibfnamefont {L.}~\bibnamefont {von
  Smekal}},\ }\href {\doibase 10.1103/PhysRevB.89.195429} {\bibfield  {journal}
  {\bibinfo  {journal} {Phys. Rev.}\ }\textbf {\bibinfo {volume} {B89}},\
  \bibinfo {pages} {195429} (\bibinfo {year} {2014})},\ \Eprint
  {http://arxiv.org/abs/1403.3620} {arXiv:1403.3620 [hep-lat]} \BibitemShut
  {NoStop}%
\bibitem [{\citenamefont {{Hohenadler}}\ \emph {et~al.}(2014)\citenamefont
  {{Hohenadler}}, \citenamefont {{Parisen Toldin}}, \citenamefont {{Herbut}},\
  and\ \citenamefont {{Assaad}}}]{2014PhRvB..90h5146H}%
  \BibitemOpen
  \bibfield  {author} {\bibinfo {author} {\bibfnamefont {M.}~\bibnamefont
  {{Hohenadler}}}, \bibinfo {author} {\bibfnamefont {F.}~\bibnamefont {{Parisen
  Toldin}}}, \bibinfo {author} {\bibfnamefont {I.~F.}\ \bibnamefont
  {{Herbut}}}, \ and\ \bibinfo {author} {\bibfnamefont {F.~F.}\ \bibnamefont
  {{Assaad}}},\ }\href {\doibase 10.1103/PhysRevB.90.085146} {\bibfield
  {journal} {\bibinfo  {journal} {Phys. Rev.}\ }\textbf {\bibinfo {volume}
  {B90}},\ \bibinfo {eid} {085146} (\bibinfo {year} {2014})},\ \Eprint
  {http://arxiv.org/abs/1407.2708} {arXiv:1407.2708 [cond-mat.str-el]}
  \BibitemShut {NoStop}%
\bibitem [{\citenamefont {Luu}\ and\ \citenamefont
  {L\"ahde}(2016)}]{Luu:2015gpl}%
  \BibitemOpen
  \bibfield  {author} {\bibinfo {author} {\bibfnamefont {T.}~\bibnamefont
  {Luu}}\ and\ \bibinfo {author} {\bibfnamefont {T.~A.}\ \bibnamefont
  {L\"ahde}},\ }\href {\doibase 10.1103/PhysRevB.93.155106} {\bibfield
  {journal} {\bibinfo  {journal} {Phys. Rev.}\ }\textbf {\bibinfo {volume}
  {B93}},\ \bibinfo {pages} {155106} (\bibinfo {year} {2016})},\ \Eprint
  {http://arxiv.org/abs/1511.04918} {arXiv:1511.04918 [cond-mat.str-el]}
  \BibitemShut {NoStop}%
\bibitem [{\citenamefont {{Braguta}}\ \emph {et~al.}(2016)\citenamefont
  {{Braguta}}, \citenamefont {{Katsnelson}}, \citenamefont {{Kotov}},\ and\
  \citenamefont {{Nikolaev}}}]{2016arXiv160807162B}%
  \BibitemOpen
  \bibfield  {author} {\bibinfo {author} {\bibfnamefont {V.~V.}\ \bibnamefont
  {{Braguta}}}, \bibinfo {author} {\bibfnamefont {M.~I.}\ \bibnamefont
  {{Katsnelson}}}, \bibinfo {author} {\bibfnamefont {A.~Y.}\ \bibnamefont
  {{Kotov}}}, \ and\ \bibinfo {author} {\bibfnamefont {A.~A.}\ \bibnamefont
  {{Nikolaev}}},\ }\href@noop {} {\bibfield  {journal} {\bibinfo  {journal}
  {ArXiv e-prints}\ } (\bibinfo {year} {2016})},\ \Eprint
  {http://arxiv.org/abs/1608.07162} {arXiv:1608.07162 [cond-mat.str-el]}
  \BibitemShut {NoStop}%
\bibitem [{\citenamefont {Hung}\ \emph {et~al.}(2014)\citenamefont {Hung},
  \citenamefont {Chua}, \citenamefont {Wang},\ and\ \citenamefont
  {Fiete}}]{Hung:2014PRB}%
  \BibitemOpen
  \bibfield  {author} {\bibinfo {author} {\bibfnamefont {H.-H.}\ \bibnamefont
  {Hung}}, \bibinfo {author} {\bibfnamefont {V.}~\bibnamefont {Chua}}, \bibinfo
  {author} {\bibfnamefont {L.}~\bibnamefont {Wang}}, \ and\ \bibinfo {author}
  {\bibfnamefont {G.~A.}\ \bibnamefont {Fiete}},\ }\href {\doibase
  10.1103/PhysRevB.89.235104} {\bibfield  {journal} {\bibinfo  {journal} {Phys.
  Rev.}\ }\textbf {\bibinfo {volume} {B89}},\ \bibinfo {pages} {235104}
  (\bibinfo {year} {2014})},\ \Eprint {http://arxiv.org/abs/1307.2659}
  {arXiv:1307.2659 [cond-mat.str-el]} \BibitemShut {NoStop}%
\bibitem [{\citenamefont {Meng}\ \emph {et~al.}(2014)\citenamefont {Meng},
  \citenamefont {Hung},\ and\ \citenamefont {Lang}}]{Meng:2014MPLB}%
  \BibitemOpen
  \bibfield  {author} {\bibinfo {author} {\bibfnamefont {Z.~Y.}\ \bibnamefont
  {Meng}}, \bibinfo {author} {\bibfnamefont {H.-H.}\ \bibnamefont {Hung}}, \
  and\ \bibinfo {author} {\bibfnamefont {T.~C.}\ \bibnamefont {Lang}},\ }\href
  {\doibase 10.1142/S0217984914300014} {\bibfield  {journal} {\bibinfo
  {journal} {Mod. Phys. Lett.}\ }\textbf {\bibinfo {volume} {B28}},\ \bibinfo
  {pages} {1430001} (\bibinfo {year} {2014})},\ \Eprint
  {http://arxiv.org/abs/1310.6064} {arXiv:1310.6064 [cond-mat.str-el]}
  \BibitemShut {NoStop}%
\bibitem [{\citenamefont {Wu}\ \emph {et~al.}(2015)\citenamefont {Wu},
  \citenamefont {He}, \citenamefont {You}, \citenamefont {Xu}, \citenamefont
  {Meng},\ and\ \citenamefont {Lu}}]{Wu:2015PRB}%
  \BibitemOpen
  \bibfield  {author} {\bibinfo {author} {\bibfnamefont {H.-Q.}\ \bibnamefont
  {Wu}}, \bibinfo {author} {\bibfnamefont {Y.-Y.}\ \bibnamefont {He}}, \bibinfo
  {author} {\bibfnamefont {Y.-Z.}\ \bibnamefont {You}}, \bibinfo {author}
  {\bibfnamefont {C.}~\bibnamefont {Xu}}, \bibinfo {author} {\bibfnamefont
  {Z.~Y.}\ \bibnamefont {Meng}}, \ and\ \bibinfo {author} {\bibfnamefont
  {Z.-Y.}\ \bibnamefont {Lu}},\ }\href {\doibase 10.1103/PhysRevB.92.165123}
  {\bibfield  {journal} {\bibinfo  {journal} {Phys. Rev.}\ }\textbf {\bibinfo
  {volume} {B92}},\ \bibinfo {pages} {165123} (\bibinfo {year} {2015})},\
  \Eprint {http://arxiv.org/abs/1506.00549} {arXiv:1506.00549
  [cond-mat.str-el]} \BibitemShut {NoStop}%
\bibitem [{\citenamefont {Yamamoto}(2016)}]{Yamamoto:2016rfr}%
  \BibitemOpen
  \bibfield  {author} {\bibinfo {author} {\bibfnamefont {A.}~\bibnamefont
  {Yamamoto}},\ }\href {\doibase 10.1103/PhysRevLett.117.052001} {\bibfield
  {journal} {\bibinfo  {journal} {Phys. Rev. Lett.}\ }\textbf {\bibinfo
  {volume} {117}},\ \bibinfo {pages} {052001} (\bibinfo {year} {2016})},\
  \Eprint {http://arxiv.org/abs/1604.08424} {arXiv:1604.08424 [hep-lat]}
  \BibitemShut {NoStop}%
\bibitem [{\citenamefont {{Chen}}\ \emph {et~al.}(2015)\citenamefont {{Chen}},
  \citenamefont {{Song}}, \citenamefont {{Jiang}}, \citenamefont {{Sun}},
  \citenamefont {{Wang}},\ and\ \citenamefont {{Xie}}}]{2015PhRvL.115x6603C}%
  \BibitemOpen
  \bibfield  {author} {\bibinfo {author} {\bibfnamefont {C.-Z.}\ \bibnamefont
  {{Chen}}}, \bibinfo {author} {\bibfnamefont {J.}~\bibnamefont {{Song}}},
  \bibinfo {author} {\bibfnamefont {H.}~\bibnamefont {{Jiang}}}, \bibinfo
  {author} {\bibfnamefont {Q.-f.}\ \bibnamefont {{Sun}}}, \bibinfo {author}
  {\bibfnamefont {Z.}~\bibnamefont {{Wang}}}, \ and\ \bibinfo {author}
  {\bibfnamefont {X.~C.}\ \bibnamefont {{Xie}}},\ }\href {\doibase
  10.1103/PhysRevLett.115.246603} {\bibfield  {journal} {\bibinfo  {journal}
  {Phys. Rev. Lett.}\ }\textbf {\bibinfo {volume} {115}},\ \bibinfo {eid}
  {246603} (\bibinfo {year} {2015})},\ \Eprint
  {http://arxiv.org/abs/1507.00128} {arXiv:1507.00128 [cond-mat.mes-hall]}
  \BibitemShut {NoStop}%
\bibitem [{\citenamefont {{Liu}}\ \emph {et~al.}(2016)\citenamefont {{Liu}},
  \citenamefont {{Ohtsuki}},\ and\ \citenamefont
  {{Shindou}}}]{2016PhRvL.116f6401L}%
  \BibitemOpen
  \bibfield  {author} {\bibinfo {author} {\bibfnamefont {S.}~\bibnamefont
  {{Liu}}}, \bibinfo {author} {\bibfnamefont {T.}~\bibnamefont {{Ohtsuki}}}, \
  and\ \bibinfo {author} {\bibfnamefont {R.}~\bibnamefont {{Shindou}}},\ }\href
  {\doibase 10.1103/PhysRevLett.116.066401} {\bibfield  {journal} {\bibinfo
  {journal} {Phys. Rev. Lett.}\ }\textbf {\bibinfo {volume} {116}},\ \bibinfo
  {eid} {066401} (\bibinfo {year} {2016})},\ \Eprint
  {http://arxiv.org/abs/1507.02381} {arXiv:1507.02381 [cond-mat.dis-nn]}
  \BibitemShut {NoStop}%
\bibitem [{\citenamefont {{Louvet}}\ \emph {et~al.}(2016)\citenamefont
  {{Louvet}}, \citenamefont {{Carpentier}},\ and\ \citenamefont
  {{Fedorenko}}}]{2016arXiv160908368L}%
  \BibitemOpen
  \bibfield  {author} {\bibinfo {author} {\bibfnamefont {T.}~\bibnamefont
  {{Louvet}}}, \bibinfo {author} {\bibfnamefont {D.}~\bibnamefont
  {{Carpentier}}}, \ and\ \bibinfo {author} {\bibfnamefont {A.~A.}\
  \bibnamefont {{Fedorenko}}},\ }\href@noop {} {\bibfield  {journal} {\bibinfo
  {journal} {ArXiv e-prints}\ } (\bibinfo {year} {2016})},\ \Eprint
  {http://arxiv.org/abs/1609.08368} {arXiv:1609.08368 [cond-mat.mes-hall]}
  \BibitemShut {NoStop}%
\bibitem [{\citenamefont {Fukui}\ \emph {et~al.}(2005)\citenamefont {Fukui},
  \citenamefont {Hatsugai},\ and\ \citenamefont {Suzuki}}]{Fukui:2005wr}%
  \BibitemOpen
  \bibfield  {author} {\bibinfo {author} {\bibfnamefont {T.}~\bibnamefont
  {Fukui}}, \bibinfo {author} {\bibfnamefont {Y.}~\bibnamefont {Hatsugai}}, \
  and\ \bibinfo {author} {\bibfnamefont {H.}~\bibnamefont {Suzuki}},\ }\href
  {\doibase 10.1143/JPSJ.74.1382, 10.1143/JPSJ.74.1674} {\bibfield  {journal}
  {\bibinfo  {journal} {J. Phys. Soc. Jap.}\ }\textbf {\bibinfo {volume}
  {74}},\ \bibinfo {pages} {1674} (\bibinfo {year} {2005})},\ \Eprint
  {http://arxiv.org/abs/cond-mat/0503172} {cond-mat/0503172
  [cond-mat.mes-hall]} \BibitemShut {NoStop}%
\bibitem [{\citenamefont {{Yoshimura}}\ \emph {et~al.}(2014)\citenamefont
  {{Yoshimura}}, \citenamefont {{Imura}}, \citenamefont {{Fukui}},\ and\
  \citenamefont {{Hatsugai}}}]{2014PhRvB..90o5443Y}%
  \BibitemOpen
  \bibfield  {author} {\bibinfo {author} {\bibfnamefont {Y.}~\bibnamefont
  {{Yoshimura}}}, \bibinfo {author} {\bibfnamefont {K.-I.}\ \bibnamefont
  {{Imura}}}, \bibinfo {author} {\bibfnamefont {T.}~\bibnamefont {{Fukui}}}, \
  and\ \bibinfo {author} {\bibfnamefont {Y.}~\bibnamefont {{Hatsugai}}},\
  }\href {\doibase 10.1103/PhysRevB.90.155443} {\bibfield  {journal} {\bibinfo
  {journal} {Phys. Rev.}\ }\textbf {\bibinfo {volume} {B90}},\ \bibinfo {eid}
  {155443} (\bibinfo {year} {2014})},\ \Eprint {http://arxiv.org/abs/1405.4842}
  {arXiv:1405.4842 [cond-mat.mes-hall]} \BibitemShut {NoStop}%
\bibitem [{\citenamefont {Rothe}(2012)}]{Rothe:1992nt}%
  \BibitemOpen
  \bibfield  {author} {\bibinfo {author} {\bibfnamefont {H.~J.}\ \bibnamefont
  {Rothe}},\ }\href {\doibase 10.1142/9789814365871_0001} {\bibfield  {journal}
  {\bibinfo  {journal} {World Sci. Lect. Notes Phys.}\ }\textbf {\bibinfo
  {volume} {82}},\ \bibinfo {pages} {1} (\bibinfo {year} {2012})}\BibitemShut
  {NoStop}%
\bibitem [{\citenamefont {Montvay}\ and\ \citenamefont
  {Munster}(1997)}]{Montvay:1994cy}%
  \BibitemOpen
  \bibfield  {author} {\bibinfo {author} {\bibfnamefont {I.}~\bibnamefont
  {Montvay}}\ and\ \bibinfo {author} {\bibfnamefont {G.}~\bibnamefont
  {Munster}},\ }\href {\doibase 10.1017/CBO9780511470783} {\emph {\bibinfo
  {title} {{Quantum fields on a lattice}}}}\ (\bibinfo  {publisher} {Cambridge
  University Press},\ \bibinfo {year} {1997})\BibitemShut {NoStop}%
\bibitem [{\citenamefont {Jansen}\ and\ \citenamefont
  {Schmaltz}(1992)}]{Jansen:1992tw}%
  \BibitemOpen
  \bibfield  {author} {\bibinfo {author} {\bibfnamefont {K.}~\bibnamefont
  {Jansen}}\ and\ \bibinfo {author} {\bibfnamefont {M.}~\bibnamefont
  {Schmaltz}},\ }\href {\doibase 10.1016/0370-2693(92)91335-7} {\bibfield
  {journal} {\bibinfo  {journal} {Phys. Lett.}\ }\textbf {\bibinfo {volume}
  {B296}},\ \bibinfo {pages} {374} (\bibinfo {year} {1992})},\ \Eprint
  {http://arxiv.org/abs/hep-lat/9209002} {arXiv:hep-lat/9209002 [hep-lat]}
  \BibitemShut {NoStop}%
\bibitem [{\citenamefont {Golterman}\ \emph {et~al.}(1993)\citenamefont
  {Golterman}, \citenamefont {Jansen},\ and\ \citenamefont
  {Kaplan}}]{Golterman:1992ub}%
  \BibitemOpen
  \bibfield  {author} {\bibinfo {author} {\bibfnamefont {M.~F.~L.}\
  \bibnamefont {Golterman}}, \bibinfo {author} {\bibfnamefont {K.}~\bibnamefont
  {Jansen}}, \ and\ \bibinfo {author} {\bibfnamefont {D.~B.}\ \bibnamefont
  {Kaplan}},\ }\href {\doibase 10.1016/0370-2693(93)90692-B} {\bibfield
  {journal} {\bibinfo  {journal} {Phys. Lett.}\ }\textbf {\bibinfo {volume}
  {B301}},\ \bibinfo {pages} {219} (\bibinfo {year} {1993})},\ \Eprint
  {http://arxiv.org/abs/hep-lat/9209003} {arXiv:hep-lat/9209003 [hep-lat]}
  \BibitemShut {NoStop}%
\bibitem [{\citenamefont {Qi}\ \emph {et~al.}(2008)\citenamefont {Qi},
  \citenamefont {Hughes},\ and\ \citenamefont {Zhang}}]{Qi:2008ew}%
  \BibitemOpen
  \bibfield  {author} {\bibinfo {author} {\bibfnamefont {X.-L.}\ \bibnamefont
  {Qi}}, \bibinfo {author} {\bibfnamefont {T.}~\bibnamefont {Hughes}}, \ and\
  \bibinfo {author} {\bibfnamefont {S.-C.}\ \bibnamefont {Zhang}},\ }\href
  {\doibase 10.1103/PhysRevB.78.195424} {\bibfield  {journal} {\bibinfo
  {journal} {Phys. Rev.}\ }\textbf {\bibinfo {volume} {B78}},\ \bibinfo {pages}
  {195424} (\bibinfo {year} {2008})},\ \Eprint {http://arxiv.org/abs/0802.3537}
  {arXiv:0802.3537 [cond-mat.mes-hall]} \BibitemShut {NoStop}%
\bibitem [{\citenamefont {Varney}\ \emph {et~al.}(2010)\citenamefont {Varney},
  \citenamefont {Sun}, \citenamefont {Rigol},\ and\ \citenamefont
  {Galitski}}]{Varney:2010PRB}%
  \BibitemOpen
  \bibfield  {author} {\bibinfo {author} {\bibfnamefont {C.~N.}\ \bibnamefont
  {Varney}}, \bibinfo {author} {\bibfnamefont {K.}~\bibnamefont {Sun}},
  \bibinfo {author} {\bibfnamefont {M.}~\bibnamefont {Rigol}}, \ and\ \bibinfo
  {author} {\bibfnamefont {V.}~\bibnamefont {Galitski}},\ }\href {\doibase
  10.1103/PhysRevB.82.115125} {\bibfield  {journal} {\bibinfo  {journal} {Phys.
  Rev.}\ }\textbf {\bibinfo {volume} {B82}},\ \bibinfo {pages} {115125}
  (\bibinfo {year} {2010})},\ \Eprint {http://arxiv.org/abs/1007.3502}
  {arXiv:1007.3502 [cond-mat.str-el]} \BibitemShut {NoStop}%
\bibitem [{\citenamefont {Varney}\ \emph {et~al.}(2011)\citenamefont {Varney},
  \citenamefont {Sun}, \citenamefont {Rigol},\ and\ \citenamefont
  {Galitski}}]{Varney:2011PRB}%
  \BibitemOpen
  \bibfield  {author} {\bibinfo {author} {\bibfnamefont {C.~N.}\ \bibnamefont
  {Varney}}, \bibinfo {author} {\bibfnamefont {K.}~\bibnamefont {Sun}},
  \bibinfo {author} {\bibfnamefont {M.}~\bibnamefont {Rigol}}, \ and\ \bibinfo
  {author} {\bibfnamefont {V.}~\bibnamefont {Galitski}},\ }\href {\doibase
  10.1103/PhysRevB.84.241105} {\bibfield  {journal} {\bibinfo  {journal} {Phys.
  Rev.}\ }\textbf {\bibinfo {volume} {B84}},\ \bibinfo {pages} {241105}
  (\bibinfo {year} {2011})},\ \Eprint {http://arxiv.org/abs/1108.2507}
  {arXiv:1108.2507 [cond-mat.str-el]} \BibitemShut {NoStop}%
\bibitem [{\citenamefont {Smit}\ and\ \citenamefont {Vink}(1988)}]{SMIT198836}%
  \BibitemOpen
  \bibfield  {author} {\bibinfo {author} {\bibfnamefont {J.}~\bibnamefont
  {Smit}}\ and\ \bibinfo {author} {\bibfnamefont {J.~C.}\ \bibnamefont
  {Vink}},\ }\href {\doibase 10.1016/0550-3213(88)90215-5} {\bibfield
  {journal} {\bibinfo  {journal} {Nucl. Phys.}\ }\textbf {\bibinfo {volume}
  {B303}},\ \bibinfo {pages} {36 } (\bibinfo {year} {1988})}\BibitemShut
  {NoStop}%
\bibitem [{\citenamefont {Vink}(1988)}]{VINK1988549}%
  \BibitemOpen
  \bibfield  {author} {\bibinfo {author} {\bibfnamefont {J.~C.}\ \bibnamefont
  {Vink}},\ }\href {\doibase 10.1016/0550-3213(88)90264-7} {\bibfield
  {journal} {\bibinfo  {journal} {Nucl. Phys.}\ }\textbf {\bibinfo {volume}
  {B307}},\ \bibinfo {pages} {549 } (\bibinfo {year} {1988})}\BibitemShut
  {NoStop}%
\bibitem [{\citenamefont {Sekine}\ and\ \citenamefont
  {Nomura}(2014)}]{Sekine:2013pda}%
  \BibitemOpen
  \bibfield  {author} {\bibinfo {author} {\bibfnamefont {A.}~\bibnamefont
  {Sekine}}\ and\ \bibinfo {author} {\bibfnamefont {K.}~\bibnamefont
  {Nomura}},\ }\href {\doibase 10.7566/JPSJ.83.094710} {\bibfield  {journal}
  {\bibinfo  {journal} {J. Phys. Soc. Jap.}\ }\textbf {\bibinfo {volume}
  {83}},\ \bibinfo {pages} {094710} (\bibinfo {year} {2014})},\ \Eprint
  {http://arxiv.org/abs/1309.1079} {arXiv:1309.1079 [cond-mat.str-el]}
  \BibitemShut {NoStop}%
\bibitem [{\citenamefont {Maciejko}\ and\ \citenamefont
  {Nandkishore}(2014)}]{Maciejko:2013lua}%
  \BibitemOpen
  \bibfield  {author} {\bibinfo {author} {\bibfnamefont {J.}~\bibnamefont
  {Maciejko}}\ and\ \bibinfo {author} {\bibfnamefont {R.}~\bibnamefont
  {Nandkishore}},\ }\href {\doibase 10.1103/PhysRevB.90.035126} {\bibfield
  {journal} {\bibinfo  {journal} {Phys. Rev.}\ }\textbf {\bibinfo {volume}
  {B90}},\ \bibinfo {pages} {035126} (\bibinfo {year} {2014})},\ \Eprint
  {http://arxiv.org/abs/1311.7133} {arXiv:1311.7133 [cond-mat.str-el]}
  \BibitemShut {NoStop}%
\bibitem [{\citenamefont {D{\'{o}}ra}\ \emph {et~al.}(2013)\citenamefont
  {D{\'{o}}ra}, \citenamefont {Herbut},\ and\ \citenamefont
  {Moessner}}]{Dora:2013PRB}%
  \BibitemOpen
  \bibfield  {author} {\bibinfo {author} {\bibfnamefont {B.}~\bibnamefont
  {D{\'{o}}ra}}, \bibinfo {author} {\bibfnamefont {I.~F.}\ \bibnamefont
  {Herbut}}, \ and\ \bibinfo {author} {\bibfnamefont {R.}~\bibnamefont
  {Moessner}},\ }\href {\doibase 10.1103/PhysRevB.88.075126} {\bibfield
  {journal} {\bibinfo  {journal} {Phys. Rev.}\ }\textbf {\bibinfo {volume}
  {B88}},\ \bibinfo {pages} {075126} (\bibinfo {year} {2013})},\ \Eprint
  {http://arxiv.org/abs/1305.7320} {arXiv:1305.7320 [cond-mat.str-el]}
  \BibitemShut {NoStop}%
\bibitem [{\citenamefont {Witczak-Krempa}\ \emph {et~al.}(2014)\citenamefont
  {Witczak-Krempa}, \citenamefont {Knap},\ and\ \citenamefont
  {Abanin}}]{Witczak-Krempa:2014nva}%
  \BibitemOpen
  \bibfield  {author} {\bibinfo {author} {\bibfnamefont {W.}~\bibnamefont
  {Witczak-Krempa}}, \bibinfo {author} {\bibfnamefont {M.}~\bibnamefont
  {Knap}}, \ and\ \bibinfo {author} {\bibfnamefont {D.}~\bibnamefont
  {Abanin}},\ }\href {\doibase 10.1103/PhysRevLett.113.136402} {\bibfield
  {journal} {\bibinfo  {journal} {Phys. Rev. Lett.}\ }\textbf {\bibinfo
  {volume} {113}},\ \bibinfo {pages} {136402} (\bibinfo {year} {2014})},\
  \Eprint {http://arxiv.org/abs/1406.0843} {arXiv:1406.0843 [cond-mat.str-el]}
  \BibitemShut {NoStop}%
\bibitem [{\citenamefont {Aoki}(1984)}]{Aoki:1983qi}%
  \BibitemOpen
  \bibfield  {author} {\bibinfo {author} {\bibfnamefont {S.}~\bibnamefont
  {Aoki}},\ }\href {\doibase 10.1103/PhysRevD.30.2653} {\bibfield  {journal}
  {\bibinfo  {journal} {Phys. Rev.}\ }\textbf {\bibinfo {volume} {D30}},\
  \bibinfo {pages} {2653} (\bibinfo {year} {1984})}\BibitemShut {NoStop}%
\bibitem [{\citenamefont {Aoki}(1986{\natexlab{a}})}]{Aoki:1985mk}%
  \BibitemOpen
  \bibfield  {author} {\bibinfo {author} {\bibfnamefont {S.}~\bibnamefont
  {Aoki}},\ }\href {\doibase 10.1103/PhysRevD.33.2399} {\bibfield  {journal}
  {\bibinfo  {journal} {Phys. Rev.}\ }\textbf {\bibinfo {volume} {D33}},\
  \bibinfo {pages} {2399} (\bibinfo {year} {1986}{\natexlab{a}})}\BibitemShut
  {NoStop}%
\bibitem [{\citenamefont {Aoki}(1986{\natexlab{b}})}]{Aoki:1986kt}%
  \BibitemOpen
  \bibfield  {author} {\bibinfo {author} {\bibfnamefont {S.}~\bibnamefont
  {Aoki}},\ }\href {\doibase 10.1103/PhysRevD.34.3170} {\bibfield  {journal}
  {\bibinfo  {journal} {Phys. Rev.}\ }\textbf {\bibinfo {volume} {D34}},\
  \bibinfo {pages} {3170} (\bibinfo {year} {1986}{\natexlab{b}})}\BibitemShut
  {NoStop}%
\bibitem [{\citenamefont {Aoki}(1986{\natexlab{c}})}]{Aoki:1986xr}%
  \BibitemOpen
  \bibfield  {author} {\bibinfo {author} {\bibfnamefont {S.}~\bibnamefont
  {Aoki}},\ }\href {\doibase 10.1103/PhysRevLett.57.3136} {\bibfield  {journal}
  {\bibinfo  {journal} {Phys. Rev. Lett.}\ }\textbf {\bibinfo {volume} {57}},\
  \bibinfo {pages} {3136} (\bibinfo {year} {1986}{\natexlab{c}})}\BibitemShut
  {NoStop}%
\bibitem [{\citenamefont {Aoki}(1989)}]{Aoki:1987us}%
  \BibitemOpen
  \bibfield  {author} {\bibinfo {author} {\bibfnamefont {S.}~\bibnamefont
  {Aoki}},\ }\href {\doibase 10.1016/0550-3213(89)90113-2} {\bibfield
  {journal} {\bibinfo  {journal} {Nucl. Phys.}\ }\textbf {\bibinfo {volume}
  {B314}},\ \bibinfo {pages} {79} (\bibinfo {year} {1989})}\BibitemShut
  {NoStop}%
\bibitem [{\citenamefont {Araki}\ and\ \citenamefont
  {Kimura}(2013)}]{Araki:2013dsa}%
  \BibitemOpen
  \bibfield  {author} {\bibinfo {author} {\bibfnamefont {Y.}~\bibnamefont
  {Araki}}\ and\ \bibinfo {author} {\bibfnamefont {T.}~\bibnamefont {Kimura}},\
  }\href {\doibase 10.1103/PhysRevB.87.205440} {\bibfield  {journal} {\bibinfo
  {journal} {Phy. Rev.}\ }\textbf {\bibinfo {volume} {B87}},\ \bibinfo {pages}
  {205440} (\bibinfo {year} {2013})},\ \Eprint {http://arxiv.org/abs/1303.1255}
  {arXiv:1303.1255 [cond-mat.str-el]} \BibitemShut {NoStop}%
\bibitem [{\citenamefont {Kimura}(2016)}]{Kimura:2015ixh}%
  \BibitemOpen
  \bibfield  {author} {\bibinfo {author} {\bibfnamefont {T.}~\bibnamefont
  {Kimura}},\ }\href
  {http://pos.sissa.it/cgi-bin/reader/conf.cgi?confid=251#session-2600}
  {\bibfield  {journal} {\bibinfo  {journal} {PoS}\ }\textbf {\bibinfo {volume}
  {LATTICE2015}},\ \bibinfo {pages} {042} (\bibinfo {year} {2016})},\ \Eprint
  {http://arxiv.org/abs/1511.08286} {arXiv:1511.08286 [cond-mat.mes-hall]}
  \BibitemShut {NoStop}%
\bibitem [{\citenamefont {Frezzotti}(2003)}]{Frezzotti:2002iv}%
  \BibitemOpen
  \bibfield  {author} {\bibinfo {author} {\bibfnamefont {R.}~\bibnamefont
  {Frezzotti}},\ }\href {\doibase 10.1016/S0920-5632(03)01502-0} {\bibfield
  {journal} {\bibinfo  {journal} {Nucl. Phys. Proc. Suppl.}\ }\textbf {\bibinfo
  {volume} {119}},\ \bibinfo {pages} {140} (\bibinfo {year} {2003})},\ \Eprint
  {http://arxiv.org/abs/hep-lat/0210007} {hep-lat/0210007 [hep-lat]}
  \BibitemShut {NoStop}%
\end{thebibliography}%

\end{document}